\documentclass[12pt]{iopart}
\pdfoutput=1

\usepackage{amstext}
\usepackage{bm}
\usepackage{graphicx}
\usepackage{subfigure}
\usepackage{cite}
\usepackage{hyperref}

\newcommand\dd{\mathrm{d}}

\newcommand\la{\lambda}
\newcommand\corr[1]{\langle #1 \rangle}
\newcommand\corrinf[1]{\langle #1 \rangle_\infty}
\newcommand\Parr{P_{\rm{arr}}}
\newcommand\Pserv{P_{\rm{serv}}}
\newcommand\Pjam{P_{\rm{jam}}}
\newcommand\corrbar[1]{\overline{ \langle #1 \rangle }}

\begin{document}

\title{Exclusion in a priority queue}
\author{$^1$Jan de Gier and ${}^2$Caley Finn}
\address{Department of Mathematics and Statistics, The University of Melbourne, 3010 VIC, Australia}
\eads{$^1$jdgier@unimelb.edu.au, $^2$c.finn3@pgrad.unimelb.edu.au}

\begin{abstract}
We introduce the prioritising exclusion process, a stochastic scheduling mechanism for a priority queueing
system in which high priority customers gain advantage by overtaking low priority customers.  The model is
analogous to a totally asymmetric exclusion process with a dynamically varying lattice length.  We calculate
exact local density profiles for an unbounded queue by  deriving domain wall dynamics from the microscopic
transition rules.  The structure of the unbounded queue carries over to bounded queues where, although no
longer exact, we find the domain wall theory is in very good agreement with simulation results.  Within this
approximation we calculate average waiting times for queueing customers.
\end{abstract}

\maketitle

\section{Introduction}

In this work we introduce the prioritising exclusion process (PEP): a stochastic scheduling mechanism for a
priority queue, where high priority customers overtake low priority customers in order to receive
service sooner.  The queue of customers is represented by a one dimensional lattice, which grows and shrinks
as customers arrive and are served.  Lattice sites are either empty or occupied by a single particle,
representing low and high priority customers respectively; particles hop forwards stochastically into empty
sites, corresponding to a high priority customer overtaking the low priority customer immediately ahead of
them, but the \textit{exclusion rule} prevents particles hopping into or over occupied sites.  The PEP is
closely related to a priority queuing model first introduced by Kleinrock \cite{Kleinrock64}, and the subject
of more recent work \cite{StanfordTZ14}.  Priority queueing systems are relevant in healthcare applications,
both as a way to efficiently manage hospital queues with patients of differing urgency \cite{StanfordTZ14},
and to describe actual practise in emergency rooms\cite{HayVB06}.

The hopping and exclusion in the PEP is analogous to that of the totally asymmetric simple exclusion process
(TASEP) \cite{Spitzer70,ASEP2}, which is one of the most thoroughly studied and central models of
non-equilibrium statistical mechanics \cite{Derrida98,Schuetz00,GolinelliM06,BlytheE07}.  The TASEP is
a microscopic model of a driven system \cite{SchmittmannZ95}, and has been the focus of much
mathematical interest due to the fact that it is integrable. Many tools have been applied to or developed for
the TASEP. Its exact stationary distribution is known \cite{DerridaDM92,SchutzD93} and can be written in
matrix product form \cite{DerridaEHP93,BlytheE07}.  Its dynamic properties are studied by means of the Bethe
ansatz \cite{GwaS92Burgers,GolinelliM06,deGierE05} and for the infinite lattice powerful techniques from
random matrix theory are available \cite{PraehoferS02}. Domain wall theory \cite{KolomeiskySKS98} provides a
phenomenological explanation of the stationary behaviour of the TASEP.  Domain wall theory is amenable to
generalisation to more complicated models that may not be integrable, and we will describe it in more detail
later.

Recently, several generalisations of the TASEP have been proposed, which allow the lattice length to vary
dynamically as is the case in the PEP.  Such models are of theoretical interest in statistical physics as they
are grand-canonical analogues of the TASEP. The review \cite{ChouMZ11} highlights many biological applications,
such as modelling filament growth \cite{SugdenEPR07, SchmittS11, DoroszMP10} and length regulation
\cite{JohannEK12,MelbingerRF12}. There have been applications to queueing theory as well.  The exclusive
queueing process (EQP) \cite{Arita09, YanagisawaTJN10} uses an exclusion process to model the motion of
customers waiting in a queue.  In the EQP, customers (all of a single priority class) are represented by
particles, with empty lattice sites for the space between them.  The hopping of particles represents customers
shuffling forwards as space becomes available.  Beyond applications, these models are also of interest because
of the rich phase structure the varying lattice length introduces \cite{SugdenE07, NowakFC07, AritaS12,AritaS13}.

As is typical of a queueing system, the PEP has a phase transition from a phase with finite expected queue
length, to one with an \textit{unbounded} queue length increasing with time.  This transition occurs when the
rate of arrival of customers exceeds the service rate.  In the latter case, we treat the lattice as infinite
in length in order to study the late time limit.  The PEP has a natural domain wall structure, and the domain
wall dynamics can be derived directly from the microscopic transition rules, similarly to \cite{CividiniHA14}.
In the unbounded phase, in the infinite lattice limit, the solution of the domain wall equations gives
exact local density profiles.  The domain wall solution reveals a second phase transition where the
`jam' of high priority customers waiting at the service end becomes infinite.

When the expected lattice length remains finite (the \textit{bounded} phase), the domain wall theory leads to
approximate solutions only, but the structure from the unbounded phase carries over.  We see a remnant of the
jamming transition from the unbounded phase as a crossover where the jam of high priority customers at the service end
delocalises, and the expected jam length becomes comparable to the queue length.  Then by defining `aggregated
correlation functions', we find that the form of the unbounded solution can be applied in the bounded phase as
an alternative to mean field theory, giving a very accurate calculation of customer waiting times.

\subsection{The model}
\begin{figure}[h]
    \centering
    \includegraphics[width=0.6\textwidth]{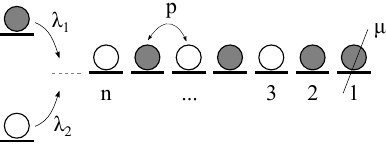}
    \caption{PEP transition rates, filled circles are occupied sites.}
    \label{fig:queuerules}
\end{figure}

In the lattice bulk, the PEP behaves as a TASEP: sites are either occupied by a single particle or empty, and
particles hop forwards into empty sites with rate $p$.  At the boundaries the PEP differs from the TASEP.  The
PEP lattice can be extended on the left by the addition of a filled or empty lattice site, with rates $\la_1$
and $\la_2$ respectively.  At the other boundary, the rightmost site is removed with rate $\mu$, irrespective
of its occupation.  These rules, summarised in Figure~\ref{fig:queuerules}, allow both the lattice length and
particle number to vary.

We specify a PEP configuration by binary variables $\tau_i$ with $\tau_i = 1$ for a filled site  and $\tau_i =
0$ for an empty site.  Usually we will number sites from right to left and write a length $n$ configuration as
\begin{equation*}
    \bm{\tau} = \tau_n\tau_{n-1}\ldots\tau_1.
\end{equation*}

\subsection{A queueing system}
\label{sec:QueueingSystem}

The PEP can be interpreted as a priority queueing system with two classes of customers.  The lattice,
itself, is the queue of customers, with filled sites representing high priority customers (class $1$) and
empty sites representing low priority customers (class $2$).  The rates $\la_1$ and $\la_2$ are the arrival
rates of high and low priority customers respectively, and the rate $\mu$, is the rate at which customers are
served and leave the queue.

In this interpretation, a particle hopping forward one site corresponds to a high priority customer stepping
ahead of the low priority customer immediately in front of them.  The stochastic overtaking is the scheduling
mechanism in this priority queue, giving high priority customers preferential treatment over low priority.
The larger the overtake rate $p$, the greater the advantage.

The PEP is modelled on a well studied priority queueing system introduced by Kleinrock
\cite{Kleinrock64,KleinrockVolII} and now known as the accumulating priority queue (APQ)
\cite{StanfordTZ14}.  In the APQ, customers have a priority value which accumulates linearly with time.  Class
$1$ customers accumulate priority faster than class $2$, thus overtaking them in the service queue.  The key
difference between the APQ and the PEP is that, for a given sequence of arrivals, overtaking in the APQ is
deterministic, but in the PEP the overtakes occur stochastically.

The PEP is also related to a simpler queueing system, the $M/M/1$ queue (see, for example,
\cite{KleinrockVolI}).  The total arrival rate of customers to the PEP is $\la = \la_1 + \la_2$, and the
service rate is $\mu$.  Both these rates are independent of the internal arrangement of the queue, and the
prioritising parameter $p$.  So, if we are interested only in the total length of the queue, we can treat the
system as a $M/M/1$ queue with arrival rate $\la$ and service rate $\mu$.  The state of a $M/M/1$ queue is
characterised simply by the length, $n$, with probability distribution $P_n$ obeying the master equation
\begin{eqnarray}
    \frac{\dd P_0}{\dd t} = \mu P_1 - \la P_0 \\
    \frac{\dd P_n}{\dd t} = \lambda P_{n-1} + \mu P_{n+1} - (\mu+\lambda)P_n, \qquad n > 0.
\end{eqnarray}
The stationary length distribution of the $M/M/1$ queue (and hence for the PEP) is the solution of
$\dd P_n / \dd t = 0$, which is
\
\begin{equation}
    P_n = \left(1 - \frac{\la}{\mu}\right)\left(\frac{\la}{\mu}\right)^n,
    \label{eq:lengthDist}
\end{equation}
when $\la < \mu$, i.e. when the total arrival rate is less than the service rate.  In this case the
system is described as \textit{stable}, because the queue length does not grow without bound.  The expected
queue length is finite, given by
\begin{equation}
    \langle n \rangle = \frac{\la}{\mu - \la}.
    \label{eq:avLength}
\end{equation}
We will call this the \textit{bounded} phase of the PEP.

When $\la > \mu$, the system is \textit{unstable} and the expected queue length grows as
\begin{equation}
    \langle n \rangle \sim (\la - \mu) t.
\end{equation}
In the late time limit, we can treat the queue as infinite in length.  We call this the \textit{unbounded}
phase of the PEP.

At the special value $p = 0$, the PEP really does reduce to a $M/M/1$ queue.  Customers arriving at rate $\la$
are high priority with probability $\la_1 / \la$ or low priority with probability $\la_2 / \la$.  But as there
is no overtaking, there is no reordering of customers, and no advantage in being a class $1$ customer.  The
probability of high or low at any site is the same as at arrival, i.e. $\la_1/\la$ or $\la_2 / \la$,
respectively.  In the bounded phase, the probability of the length $n$ configuration $\tau_n\ldots\tau_1$ is
then
\begin{equation}
    P(\tau_n\ldots\tau_1) = P_n \left(\frac{\la_1}{\la}\right)^h \left(\frac{\la_2}{\la}\right)^l,
    \label{eq:pZeroExact}
\end{equation}
where $h$ is the number of high priority customers, and $l$ the number of low priority customers, i.e.
\begin{equation}
    h = \sum_{i=1}^n \tau_i, \qquad l = n - h.
\end{equation}

We can contrast the phase behaviour of the PEP with that of the EQP, where the length of the lattice is
defined by the position of the last customer, and so the lattice length depends on how fast customers step
into the space ahead of them (i.e. the particle hopping rate).  The EQP has bounded and unbounded length
phases\footnote{convergent and divergent in their terminology}, but with phase boundaries dependent on the
hopping rate \cite{Arita09}.

\subsection{Density profiles and waiting times}

To define a density profile for the PEP we must specify both the site, $i$, and the lattice length, $n$, so
that
\begin{equation*}
    \corr{\tau_i}_n = P(\text{queue length is } n \text{, and site } i \text{ is occupied}).
\end{equation*}
These are one-point functions.  We can similarly define higher order correlations
\begin{equation}
    \corr{\tau_{i_1} \tau_{i_2} \ldots \tau_{i_m}}_n, \qquad n \ge i_1 > i_2 > \ldots > i_m \ge 1.
\end{equation}

The rate equations for the one-point functions are
\begin{eqnarray}
    \fl \frac{\dd}{\dd t} \corr{\tau_1}_1 = \la_1 P_0 + \mu \corr{\tau_2}_2
                                          - (\la + \mu) \corr{\tau_1}_1, 
    \label{eq:rate11} \\
    \fl \frac{\dd}{\dd t} \corr{\tau_1}_n = \la \corr{\tau_1}_{n-1} + \mu \corr{\tau_2}_{n+1}
                                          + p \corr{\tau_2(1 - \tau_1)}_n
                                          - (\la + \mu)\corr{\tau_1}_n, \qquad n > 1,
    \label{eq:rate1n} \\
    \fl \frac{\dd}{\dd t} \corr{\tau_i}_i = \la_1 P_{i-1} + \mu \corr{\tau_{i+1}}_{i+1}
                                -p \corr{\tau_i(1 - \tau_{i-1})}_i - (\la + \mu) \corr{\tau_i}_i, \qquad i > 1,
    \label{eq:rateii} \\
    \fl \frac{\dd}{\dd t} \corr{\tau_i}_n = \la \corr{\tau_i}_{n-1} + \mu \corr{\tau_{i+1}}_{n+1}
                                + p \corr{\tau_{i+1}(1 - \tau_i)}_n                     \nonumber \\
                             {} - p \corr{\tau_i(1 - \tau_{i-1})}_n - (\la + \mu) \corr{\tau_i}_n, \qquad i > 1, n > i.
    \label{eq:ratein}
\end{eqnarray}
These couple the one-point functions to the two-point correlations, and length $n$ to length $n \pm 1$.  The
rate equations imply a conserved current of particles across the lattice, but because of the coupling between
lengths, some care is required in how this current is defined.  We will return to this for the bounded and
unbounded phases separately.

Viewing the PEP as a queueing system, we are interested in performance measures, and how these differ for high
and low priority customers.  The current tells us the rate at which customers pass through the system, and
from the density profile we can calculate the average waiting time for customers of each class.

To calculate waiting times, we use Little's result (see Chapter 2.1 of \cite{KleinrockVolI}), which states
that the average waiting time, $\overline{W}_i$, is related to the average number of waiting customers,
$\overline{N}_i$, for each class $i = 1, 2$, by
\begin{equation}
    \overline{N}_i = \la_i \overline{W}_i.
    \label{eq:littles}
\end{equation}
The average number of high priority customers can be calculated from the density profile as
\begin{equation}
    \overline{N}_1 = \sum_{n=1}^\infty \sum_{i=1}^n \corr{\tau_i}_n,
    \label{eq:pepNhigh}
\end{equation}
and the average number of low priority customers is
\begin{equation}
    \overline{N}_2 = \corr{n} - \overline{N}_1 = \frac{\la}{\mu - \la} - \overline{N}_1.
    \label{eq:pepNlow}
\end{equation}
Here we take the total time from arrival to removal from the system as the waiting time for a customer.  Our
aim, now, is to compute the density profile for the PEP.

\subsection{Domain wall theory}
\label{sec:DWT}

Domain wall theory \cite{KolomeiskySKS98, SantenA02} reduces the multi-particle dynamics of the
TASEP\footnote{Domain wall theory applies more generally to the \textit{partially} asymmetric simple exclusion
process.} to the
motion of a single random walker on the lattice.  The TASEP boundary conditions (the particle entry and exit
rates) create domains of low or high density at the boundaries.  These domains extend through the lattice, and
where they  meet a shock, or domain wall, forms.  Domain wall theory models the motion of this shock as a
random walk, with the simplifying assumptions that density is constant throughout each domain, and that there
is a sharp transition between domains so that the shock can be localised to a single site.  Though the
stationary solution of the TASEP is known exactly, domain wall theory provides a simple physical explanation
of the stationary behaviour \cite{KolomeiskySKS98}, and beyond this it allows accurate approximation of
some dynamic properties \cite{SantenA02}.  Domain wall theory can also be applied to more complex models
\cite{PopkovSSS01,CookZ09} where the exact solution is not known.  For the EQP, a domain wall approach was used to describe the global density profile in the divergent length
(i.e. unbounded) phase \cite{AritaS12}. In \cite{CividiniHA14} domain wall theory provided an exact solution and we will show that this also occurs in the unbounded phase of the PEP.

\begin{figure}
    \centering
    \subfigure[A queue configuration with jam length $k = 3$]
    {
        \includegraphics[width=0.4\textwidth]{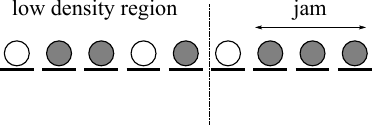}
        \label{fig:jam}
    }
    \subfigure[Conditional probability that site $i$ is filled given queue length $n$ and jam length $k$]
    {
        \includegraphics[width=0.4\textwidth]{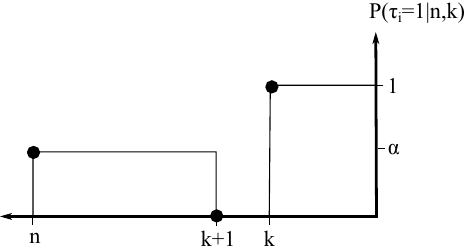}
        \label{fig:dwprob}
    }
    \caption{}
\end{figure}

The PEP has a natural domain wall structure.  As high priority customers overtake and reach the service end,
they form a jam (Figure \ref{fig:jam}): a jam is a section of high priority customers (filled sites) at the
service end ahead of any low priority customer (empty site).  The jam is characterised by $k$, the number of
consecutive high priority customers.  As there are no gaps, there is no overtaking in the jammed region, and
the length of the jam reduces only as customers are served.  This is similar to the situation in
\cite{CividiniHA14}, where a TASEP with parallel update and deterministic bulk motion is considered.  In
\cite{deGierFS11}, a jam of particles was suggested as the cause of a reduced effective lattice length in the
reverse bias regime of the partially asymmetric simple exclusion process.

Let us assume that the region beyond the jam has uniform density, and that the conditional probability that site
$i$ is filled, given the queue length, $n$, and jam length, $k$, is (Figure~\ref{fig:dwprob})
\begin{equation}
    P(\tau_i = 1 | n, k) =
    \cases{
        1      & $1 \le i \le k$ \\
        0      & $i = k + 1$ \\
        \alpha & $k + 2 \le i \le n$.
    }
    \label{eq:dwprob}
\end{equation}
Then the bulk equation ($n > k+1, k > 0$) for $P(n,k)$, the probability of a length $k$ jam in a length $n$
queue, is
\begin{equation}
    \eqalign{
\fl     \frac{\dd}{\dd t} P(n, k) = & \la P(n - 1, k) + \mu P(n + 1, k + 1) + \mu (1 - \alpha) \alpha^k P(n + 1, 0) \\
                                    &  {} + p \alpha P(n, k - 1) - (\la + \mu + p \alpha) P(n, k).
    }
    \label{eq:effDW}
\end{equation}

Let us explain the meaning of each of the terms in equation \eref{eq:effDW}.  The term
\begin{equation*}
    \la P(n - 1, k)
\end{equation*}
is the entry into the $(n, k)$ configuration from a length $n$ queue due to the arrival of a customer,
and the terms
\begin{equation*}
    \mu P(n + 1, k + 1) + \mu (1 - \alpha) \alpha^k P(n + 1, 0),
\end{equation*}
represent the service of a customer.  The second term is the transition into the $k$-jam state from the $(k =
0)$-jam state  by the service of a low priority customer who was followed by $k$ consecutive high priority
customers.  The term
\begin{equation*}
    p \alpha P(n, k - 1),
\end{equation*}
is a $(k-1)$-jam extending to length $k$ with rate $p \alpha$: there is a high priority customer at site $k +
1$ with probability $\alpha$, which overtakes with rate $p$ the low priority customer at site $k$ (the low
priority customer marking the end of the $(k-1)$-jam).  The low priority customer thus moves to position $k+1$
defining the new end of the jam.  The loss terms
\begin{equation*}
    -(\la + \mu + p \alpha) P(n, k),
\end{equation*}
are the rate at which the $(n, k)$ configuration is left due to a customer arrival or service, or growth of the jam.

In the next section, we will show that the $n \to \infty$ limit of \eref{eq:effDW}, and the corresponding
equation for $k = 0$, follows from the unbounded phase master equation.  We find the exact stationary
solutions of these equations, describing the behaviour of the jam on an infinite lattice.  In the bounded
phase (Section~\ref{sec:bounded_phase}), domain wall theory leads to two complementary approximations. One reveals information about the length dependence and the other about waiting times.

\section{The unbounded phase}

We consider first the unbounded phase of the PEP, where the total arrival rate exceeds the service rate ($\la >
\mu$).  Recall that the expected lattice length grows as $\corr{n} \sim (\la - \mu)t$.  In our domain wall
picture, the jam increases with rate $p \alpha$ and decreases with rate $\mu$ (ignoring for the moment the
$\mu(1 - \alpha)\alpha^k P(n + 1, 0)$ term in \eref{eq:effDW}).  If $p \alpha > \mu$, the jam will grow as
$(p \alpha - \mu) t$ unless it reaches the arrival end of the queue, but if $p \alpha < \mu$, the jam length
will fluctuate near $0$.

\begin{figure}[ht]
    \centering
    \subfigure[
        With $p = 3$, there is a growing jam.
    ]
    {
        \includegraphics[width=0.45\textwidth]{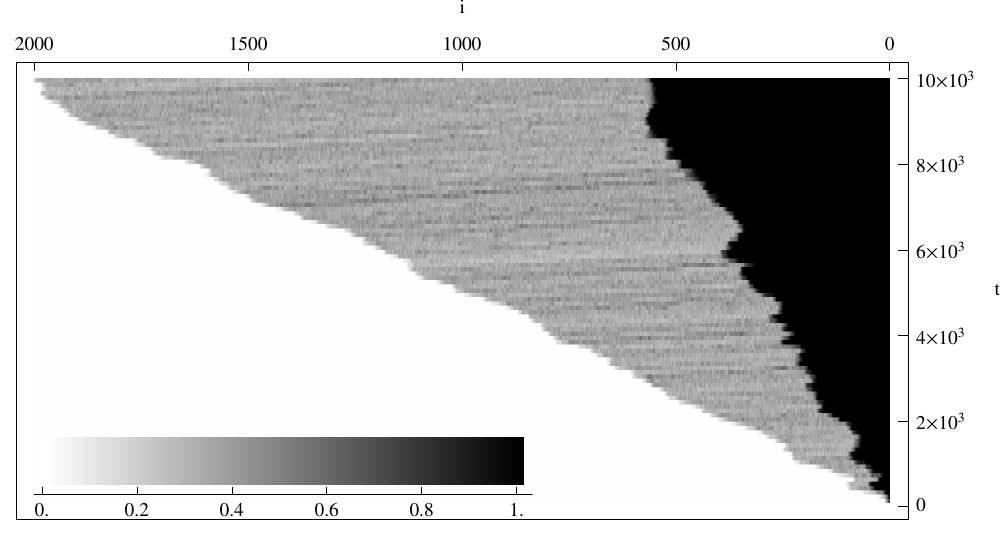}
        \label{fig:unbounded_unbounded_time}
    }
    \qquad
    \subfigure[
        With $p = 1.8$ the jam remains finite.
    ]
    {
        \includegraphics[width=0.45\textwidth]{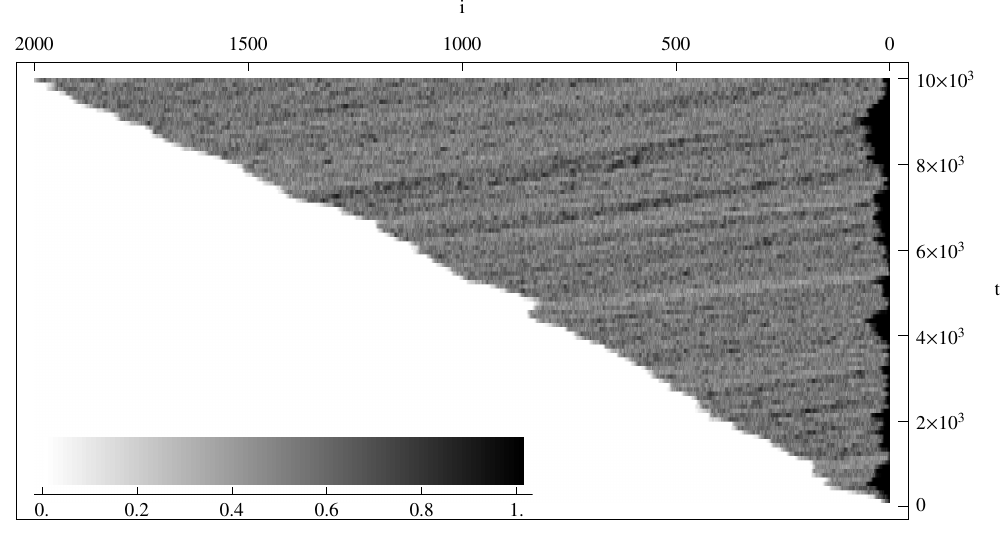}
        \label{fig:unbounded_bounded_time}
    }
    \caption{
        Time evolution of an unbounded queue with arrival rates $\la_1 = 1.1$, $\la_2 = 0.1$,
        and service rate $\mu = 1$.  Density $\corr{\tau_i}(t)$ is averaged over a small time period,
        and the inset shows mapping of density to colour.
    }
    \label{fig:unbounded_evolution}
\end{figure}

Figure~\ref{fig:unbounded_evolution} shows simulation results for a growing jam
(Figure~\ref{fig:unbounded_unbounded_time}), and a jam fluctuating near $0$
(Figure~\ref{fig:unbounded_bounded_time}).  The figures show the time evolution of the density profile,
starting from an empty queue, with the density at each site calculated by averaging over a short time period.
The queue length grows with rate $\la - \mu$, and the low density region beyond the jam of high priority
customers is fairly regular.  We will focus on the situation in Figure~\ref{fig:unbounded_bounded_time}, where
the jam length remains finite.  In
Figure~\ref{fig:unbounded_unbounded_time} where the jam grows, we see that it nevertheless grows more slowly
than the queue length, and we will comment on this later.

We would like to understand the late time behaviour in the unbounded phase, but as for any fixed configuration
$\tau_n\ldots\tau_1$,
\begin{equation}
    \lim_{t \to \infty} P(\tau_n \ldots \tau_1) = 0,
\end{equation}
the \textit{global} description is uninformative.  Instead, we follow the approach of \cite{SugdenE07} and
consider \textit{local} behaviour relative to a specified reference frame.

The \textit{service frame} is fixed at the right hand end of the lattice where customers are served and
depart.  For a general but finite section of length $m$,
\begin{equation*}
    \tau_m\ldots\tau_1,
\end{equation*}
the service frame probability in a length $n$ lattice is defined as
\begin{equation}
    \Pserv(\tau_m\ldots\tau_1; n) = \sum_{\tau_n,\ldots,\tau_{m+1}=0,1}
                                         P(\tau_n\ldots\tau_{m+1}\tau_m\ldots\tau_1).
\end{equation}
We will also consider the \textit{arrival frame}, fixed at the left hand end of the lattice.  In this case we
number sites left to right.  A general length $m$ section is written
\begin{equation*}
    \tau_1\ldots\tau_m
\end{equation*}
and the arrival frame probability in a length $n$ lattice is defined by
\begin{equation}
    \Parr(\tau_1\ldots\tau_m; n) = \sum_{\tau_{m+1},\ldots,\tau_n=0,1}
                                        P(\tau_1\ldots\tau_{m}\tau_{m+1}\ldots\tau_n).
\end{equation}

In the $t \to \infty$ limit, the expected lattice length is infinite, and thus we are interested in the
$n \to \infty$ limits
\begin{equation}
\eqalign{
    \Pserv(\tau_m\ldots\tau_1) = \lim_{n \to \infty} \Pserv(\tau_m\ldots\tau_1; n), \\
    \Parr(\tau_1\ldots\tau_m) = \lim_{n \to \infty} \Parr(\tau_1\ldots\tau_m; n), \\
}
\label{eq:ub:localProbs}
\end{equation}
with the assumption that this limit exists.  We will write the rate equations for the arrival and service
frame probabilities \eref{eq:ub:localProbs} in this limit, then seek the stationary solution.  To do this, we
use a domain wall ansatz for \eref{eq:ub:localProbs}, and show that this leads to the exact stationary solution
for these quantities.

\subsection{Domain wall ansatz}
We use domain wall theory to form an ansatz for the service frame probabilities.  We consider a general but
finite section of length $m$, with a length $k$ jam,
\begin{equation}
    \tau_m\ldots\tau_{k+2}01^k = \bm{\tau}01^k,
    \label{eq:ub:kjamconfig}
\end{equation}
where $1^k$ indicates a string of $k$ $1$'s.  The configuration of any finite segment can be written this way,
as long as we can take $m \ge k + 1$.  We will assume that the conditional probability for a high at site $i$
given a jam of length $k$ is
\begin{equation}
    P(\tau_i = 1 | k) =
    \cases{
        1      & $1 \le i \le k$ \\
        0      & $i = k + 1$ \\
        \alpha & $k + 2 \le i$,
    }
\end{equation}
which is \eref{eq:dwprob} in the $n \to \infty$ limit.  Then the probability of the finite segment
\eref{eq:ub:kjamconfig} is
\begin{equation}
    \eqalign{
        \Pserv(\bm{\tau}01^k) &= \sum_{\tau_\infty,\ldots,\tau_{m+1}=0,1}P(\ldots\tau_{m+1}\tau_m\ldots\tau_{k+2}01^k) \\
                 &= \alpha^h (1 - \alpha)^l \Pjam(k),
    }
    \label{eq:ub:jamAnsatz}
\end{equation}
where $h$ is the number of highs in the configuration beyond the jam up to position $m$, and $l$ is the number
of lows, that is
\begin{equation}
    h = \sum_{i = k+2}^m \tau_i, \qquad l = m - h - k - 1,
\end{equation}
and $\Pjam(k)$ is the probability of a length $k$ jam.  The jam probabilities are normalised such that
\begin{equation}
    \sum_{k = 0}^\infty \Pjam(k) = 1.
    \label{eq:ub:jamNorm}
\end{equation}
Equation \eref{eq:ub:jamAnsatz} is the domain wall ansatz for the stationary service frame probabilities.

\subsection{The service frame}
\label{sec:ub:service_frame}

In this section we write the general service frame rate equations, then apply the domain wall ansatz
\eref{eq:ub:jamAnsatz}.  We use the notation $\bm{\tau}|_{(i,i-1)}$ to indicate the exchange of customers in
places $i$ and $i - 1$. That is, for $\bm{\tau} = \tau_r\ldots\tau_1$
\begin{equation*}
    \bm{\tau}|_{(i,i-1)} = \tau_r\ldots\tau_{i+1}\tau_{i-1}\tau_i\tau_{i-2}\ldots\tau_1,
\end{equation*}
and
\begin{equation*}
    0\bm{\tau}|_{(r+1,r)} = \tau_r 0 \tau_{r-1}\ldots\tau_1.
\end{equation*}

The stationary rate equation for the $k$-jam configuration \eref{eq:ub:kjamconfig} with $k \geq 1$ is
\begin{eqnarray}
    \fl 0 &= \frac{\dd}{\dd t} \Pserv(\bm{\tau}01^k) \nonumber \\
      \fl &= \mu \Pserv\left(\bm{\tau}01^{k+1}\right) + \mu \Pserv\left(\bm{\tau}01^k0\right) \nonumber \\
        \fl & {} + p \tau_m \Pserv\left(0\bm{\tau}01^k|_{(m+1,m)}\right)
      + \sum_{i = k + 2}^m p (1 - \tau_i)\tau_{i-1} \Pserv\left(\bm{\tau}01^k|_{(i,i-1)}\right)
           + p \Pserv\left(\bm{\tau}101^{k-1}\right) \nonumber \\
      \fl & {} -\mu \Pserv(\bm{\tau}01^k) - \sum_{i = k + 2}^m p \tau_i (1 - \tau_{i-1}) \Pserv(\bm{\tau}01^k)
           - p(1 - \tau_m) \Pserv(1\bm{\tau}01^k).
           \label{eq:ub:statrate}
\end{eqnarray}
Let us again explain the various terms. The terms
\begin{equation*}
    \mu \Pserv\left(\bm{\tau}01^{k+1}\right) + \mu \Pserv\left(\bm{\tau}01^k0\right),
\end{equation*}
give the rate of arrival to the $k$-jam configuration after, respectively, a high or low priority
customer is served. Then there are the hopping terms.  A high in $m$th place in $\bm{\tau}01^k$ can arrive
from place $m + 1$:
\begin{equation*}
    p \tau_m \Pserv\left(0\bm{\tau}01^k|_{(m+1,m)}\right).
\end{equation*}
Overtaking within the low density region behind the jam is given by
\begin{equation*}
    \sum_{i = k + 2}^m p (1 - \tau_i)\tau_{i-1} \Pserv\left(\bm{\tau}01^k|_{(i,i-1)}\right),
\end{equation*}
and a $(k-1)$-jam extends to a $k$-jam when a high hops onto the end:
\begin{equation*}
    p \Pserv\left(\bm{\tau}101^{k-1}\right).
\end{equation*}
The loss term
\begin{equation*}
    -\mu \Pserv(\bm{\tau}01^k),
\end{equation*}
is the reduction of the jam as a customer is served, and
\begin{equation*}
    -\sum_{i = k + 2}^m p \tau_i (1 - \tau_{i-1}) \Pserv(\bm{\tau}01^k)
\end{equation*}
are overtakings within the $m$ places of $\bm{\tau}01^k$.  The final loss term
\begin{equation*}
    - p(1 - \tau_m) \Pserv(1\bm{\tau}01^k)
\end{equation*}
arises if the configuration has a low in $m$th place, which can be overtaken by a high from place $m+1$. Finally we note that the terms involving the arrival rates $\lambda_1$ and $\lambda_2$ do not appear in \eref{eq:ub:statrate} as they cancel from the stationary rate equations in the $n\rightarrow\infty$ limit.

Substituting the ansatz \eref{eq:ub:jamAnsatz}, the terms representing overtaking within the $m$ sites
of $\bm{\tau}01^k$ combine and telescope to
\begin{eqnarray*}
     p \left(\sum_{i=k+2}^m (1 - \tau_i)\tau_{i-1} - \sum_{i=k+2}^m \tau_i(1 - \tau_{i-1})\right)
        \alpha^h (1 - \alpha)^l \Pjam(k) \\
        = p \left(\tau_{k+1} - \tau_m\right)\alpha^h (1 - \alpha)^l \Pjam(k) \\
        = - p \tau_m\alpha^h (1 - \alpha)^l \Pjam(k);
\end{eqnarray*}
recall that $\tau_{k+1} = 0$.

The factor $\alpha^h (1 - \alpha)^l$ is common to all terms in  the rate equation.  Cancelling, and
simplifying leaves
\begin{equation}
\fl  0 = p \alpha \Pjam(k-1) + \mu \Pjam(k+1) + \mu (1 - \alpha)\alpha^k \Pjam(0) - (\mu + p \alpha) \Pjam(k).
    \label{eq:ub:exactDW}
\end{equation}
This agrees with the $n \to \infty$ limit of \eref{eq:effDW}, but we have derived it from the
full PEP rate equations.  Were it not for the $\Pjam(0)$ term, this equation for the position of the jam
would have the same form as the domain wall theory for the TASEP \cite{SantenA02}.

The $k = 0$ case differs only slightly. In this case the rate equation is
\begin{eqnarray}
     \fl 0 &= \frac{\dd}{\dd t} \Pserv(\bm{\tau}0) \nonumber \\
     \fl   &= \mu \Pserv\left(\bm{\tau}01\right) + \mu \Pserv\left(\bm{\tau}00\right) \nonumber \\
     \fl   & {} + p \tau_m \Pserv\left(0\bm{\tau}0|_{(m+1,m)}\right)
                + \sum_{i = 2}^m p (1 - \tau_i)\tau_{i-1} \Pserv\left(\bm{\tau}0|_{(i,i-1)}\right) \\
     \fl   & {} - \mu P\left(\bm{\tau}0\right) - \sum_{i = 2}^m p \tau_i (1 - \tau_{i-1})  \Pserv\left(\bm{\tau}0\right)
                - p(1 - \tau_m) \Pserv\left(1\bm{\tau}0\right) \nonumber
\end{eqnarray}
Substituting the ansatz \eref{eq:ub:jamAnsatz}, this reduces to
\begin{equation}
    0 = \mu \Pjam(1) - (\mu \alpha + p \alpha) \Pjam(0),
    \label{eq:ub:zeroDW}
\end{equation}
and rearranging gives
\begin{equation}
    \Pjam(1) = \frac{p \alpha}{\mu} \Pjam(0) + \alpha \Pjam(0).
\end{equation}
With this as the base case, we use \eref{eq:ub:exactDW} to show by induction that
\begin{equation}
    \Pjam(k) = \frac{p \alpha}{\mu} \Pjam(k-1) + \alpha^k \Pjam(0), \qquad k \ge 1.
\end{equation}
This recurrence for $\Pjam(k)$ has solution
\begin{eqnarray}
    \Pjam(k) &= \sum_{i=0}^k \left(\frac{p \alpha}{\mu}\right)^{k-i} \alpha^i \Pjam(0) \nonumber \\
             &= \frac{p \left(\frac{p \alpha}{\mu}\right)^k - \mu \alpha^k}{p - \mu} \Pjam(0).
    \label{eq:ub:PjamSol}
\end{eqnarray}
The normalisation condition \eref{eq:ub:jamNorm} fixes
\begin{equation}
    \Pjam(0) = (1 - \alpha)(1 - \frac{p \alpha}{\mu}),
    \label{eq:ub:PjamNorm}
\end{equation}
subject to the constraint
\begin{equation}
    p \alpha < \mu.
    \label{eq:ub:jamConstraint}
\end{equation}

The domain wall picture makes the meaning of this constraint clear. The jam of high
priority customers grows with rate $p \alpha$ and is reduced with rate $\mu$.  If $p \alpha > \mu$ the jam
grows with rate $p \alpha - \mu > 0$, that is
\begin{equation}
    \langle k \rangle \sim (p \alpha - \mu) t,
\end{equation}
and as $t \to \infty$ the expected length of the jam becomes infinite.  In contrast, when
\eref{eq:ub:jamConstraint} is satisfied, the service rate is fast enough to prevent a backlog of high
priority customers, and the expected jam length is finite and given by
\begin{equation}
    \langle k \rangle = \sum_{k=1}^\infty k \Pjam(k)
        = \frac{\alpha}{1 - \alpha} + \frac{p \alpha}{\mu - p \alpha}.
\end{equation}

\subsection{The arrival frame}
To determine $\alpha$, we examine the PEP in the arrival frame.  Sites are now numbered left to right, and the
interchange operation is defined as
\begin{equation*}
    \bm{\tau}|_{(i,i+1)} = \tau_1\ldots\tau_{i-1}\tau_{i+1}\tau_i\tau_{i+2}\ldots\tau_r,
\end{equation*}
for $\bm{\tau} = \tau_1\ldots\tau_r$.

We will assume that the jam is always far from the arrival end.  In the late time limit this is guaranteed if
\begin{equation}
    p \alpha < \la.
    \label{eq:ub:arrivalConstraint}
\end{equation}
This condition is clearly met when \eref{eq:ub:jamConstraint} is satisfied, but we will show that $\alpha$
can be determined consistently with this requirement.  Then in the arrival frame, the ansatz
\eref{eq:ub:jamAnsatz} implies that for a configuration on the first $m$ sites,
\begin{equation*}
    \bm{\tau} = \tau_1\ldots\tau_m,
\end{equation*}
the arrival frame probability has the form
\begin{equation}
    \Parr(\bm{\tau}) = \alpha^h(1 - \alpha)^l,
    \label{eq:ub:arrAnsatz}
\end{equation}
where
\begin{equation*}
    h = \sum_{i = 1}^m \tau_i, \qquad l = m - h.
\end{equation*}

The stationary rate equation for this configuration is
\begin{eqnarray}
    0 &= \frac{\dd}{\dd t}\Parr(\bm{\tau}) \nonumber \\
      &= \la_1 \tau_1 \Parr(\tau_2\ldots\tau_m) + \la_2 (1 - \tau_1) \Parr(\tau_2\ldots\tau_m) \nonumber \\
      & {} + \sum_{i=1}^{m-1} p (1 - \tau_i)\tau_{i+1} \Parr(\bm{\tau}|_{(i,i+1)})
         + p (1 - \tau_m) \Parr(\bm{\tau}1|_{(m,m+1)}) \\
      & {} -\la \Parr(\bm{\tau}) - \sum_{i = 1}^{m-1} p \tau_i(1 - \tau_{i+1}) \Parr(\bm{\tau})
             - p\tau_m \Parr(\bm{\tau}0). \nonumber
\end{eqnarray}
Substituting \eref{eq:ub:arrAnsatz}, the summed hopping terms again combine and telescope, and the
factors of $\alpha$ and $1 - \alpha$ common to all terms can be cancelled.  This leaves
\begin{equation*}
    0 = -\la \alpha(1 - \alpha) + p \alpha^2(1 - \alpha) - p \tau_1 \alpha (1 - \alpha)
        + \tau_1 \la_1(1 - \alpha) + (1 - \tau_1) \la_2 \alpha,
\end{equation*}
which for both $\tau_1 = 0$ and $\tau_1 = 1$ reduces to
\begin{equation*}
    p \alpha^2 - (p + \la) \alpha + \la_1 = 0.
\end{equation*}
The two solutions are
\begin{equation}
    \alpha_\pm = \frac{p + \la \pm \sqrt{(p - \la)^2 + 4 p \la_2}}{2p},
    \label{eq:ub:alphapm}
\end{equation}
with $0 < \alpha_- < 1$ and $\alpha_+ > 1$ for $p, \la_1, \la_2 > 0$.  As $\alpha$ is a density value we
must take $\alpha = \alpha_-$.  Substituting this value for $\alpha$, it is seen that $p \alpha < \la$ (the
constraint \eref{eq:ub:arrivalConstraint}) is satisfied for all physical parameter values.  The jam always
grows more slowly than the queue length, as illustrated by the simulation results in
Figure~\ref{fig:unbounded_unbounded_time}.  In the limit $p \to 0$, $\alpha \to \la_1 / \la$, the expected
occupancy of any site when there is no overtaking.

With $\alpha$ given by \eref{eq:ub:alphapm}, and the jam probabilities, $\Pjam(k)$ by \eref{eq:ub:PjamSol} and
\eref{eq:ub:PjamNorm}, we thus have the stationary domain wall solution describing the motion of the high
priority jam on an infinite lattice.  Through \eref{eq:ub:jamAnsatz} and \eref{eq:ub:arrAnsatz}, this then
gives the exact stationary service and arrival frame probabilities $\Pserv(\tau_m\ldots\tau_1)$ and
$\Parr(\tau_1\ldots\tau_m)$.  In the next section, we compare this solution to simulation results, where the
lattice length, $n$,  is large, but necessarily finite.  In doing so we make the assumption that the large $n$
behaviour converges to the $n \to \infty$ limit.

We note also that the result
\eref{eq:ub:PjamNorm}, the probability to not have a jam, can be understood by the following heuristic
argument\footnote{We thank an anonymous referee for pointing this out.}. The number of low priority customers
in the system at time $t$ can be expressed as $(1-\alpha)(\lambda- \mu)t$, i.e. the ratio of low priority
customers $(1-\alpha)$ times the expected length of the queue. It can also be expressed as $(\lambda_2-P_{\rm
jam}(0)\mu)t$, i.e. (arrival rate $-$ service rate) times time. Comparing the two expressions and using
\eref{eq:ub:alphapm} we obtain $\eref{eq:ub:PjamNorm}$.

\subsection{Density profile, conserved currents, and service rates}

\begin{figure}[t]
    \centering
    \subfigure[Density profile at the service end of the unbounded queue; 
        $\la_1 = 1.1, \la_2 = 0.1, \mu = 1$ with $p = 1$ (black), $p = 1.4$ (mid-gray), and
        $p = 1.8$ (light gray).
    ]
    {
        \includegraphics[width=0.4\textwidth]{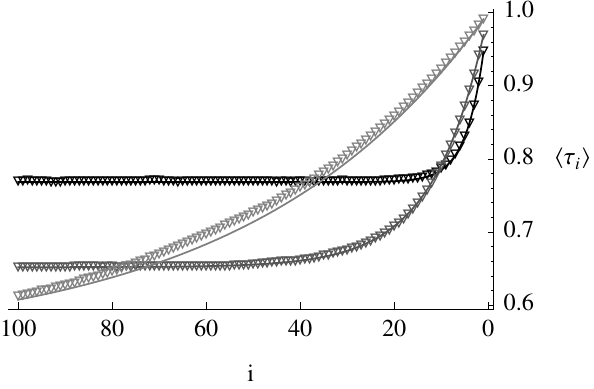}
        \label{fig:unbounded_density_a}
    }
    \subfigure[Density profile at the service end of the unbounded queue;
        $\la_1 = 0.9, \la_2 = 0.3, \mu = 1$ with $p = 1$ (black), $p = 1.8$ (mid-gray), and
        $p = 3$ (light gray).
    ]
    {
        \includegraphics[width=0.4\textwidth]{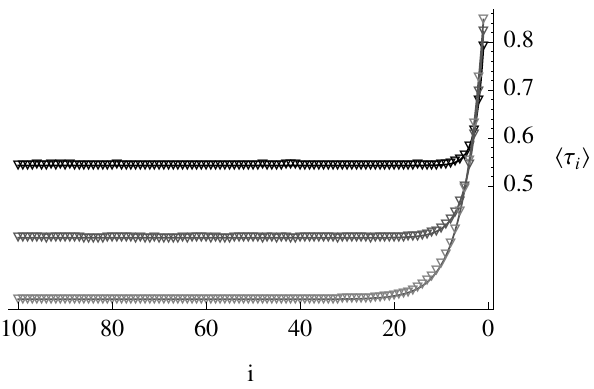}
        \label{fig:unbounded_density_b}
    }
    \caption{
        Density profiles of the unbounded PEP in the service frame.  Triangle markers for simulation
        results plotted against calculated profile $\corr{\tau_i}$.
    }
    \label{fig:unbounded_density}
\end{figure}

The density at site $i$ in the service frame, $\corrinf{\tau_i}$, is computed from the domain wall solution as
\begin{eqnarray}
    \corrinf{\tau_i} &= \alpha \sum_{k = 0}^{i - 2}\Pjam(k) + \sum_{k = i}^\infty\Pjam(k) \nonumber \\
                  &= \alpha + (1 - \alpha)\left(\frac{p \alpha}{\mu}\right)^i.
    \label{eq:ub:onePoint}
\end{eqnarray}
Figure~\ref{fig:unbounded_density} shows $\corrinf{\tau_i}$ computed from \eref{eq:ub:onePoint} plotted
against simulation results for a range of parameters chosen, necessarily, with $p \alpha < \mu$.  In all cases
we see excellent agreement.  The only deviation occurs for $p = 1.8$ in Figure~\ref{fig:unbounded_density_a},
which with $p \alpha \simeq 0.98$ is very close to the critical value $p \alpha = \mu$.  We believe that this
difference occurs because of the slowing convergence of simulation results as we near the critical point.
A study of the critical behaviour, as for example the numerical study\cite{AritaS13} for the EQP, would be
required to confirm this.
With $p \alpha > \mu$, the jam grows to fill any finite section at the service end.  Checking
Figure~\ref{fig:unbounded_unbounded_time}, we see that the rate of growth is consistent with
$\corr{k} \sim (p \alpha - \mu) t$.

We can check that the domain wall solution satisfies the rate equations for the one point functions
\eref{eq:rate11} -- \eref{eq:ratein}.  To obtain the service frame rate equations, we take the $n \to \infty$
limit, allowing us to neglect the boundary cases.  The bulk equations can be written
\begin{equation}
    \frac{\dd}{\dd t} \corrinf{\tau_i} = J_\infty^{(i+1)} - J_\infty^{(i)}, \qquad i \ge 1,
    \label{eq:ub:dtJ}
\end{equation}
where
\begin{eqnarray}
    J_\infty^{(1)} = \mu \corrinf{\tau_1} \nonumber \\
    J_\infty^{(i)} = \mu \corrinf{\tau_{i}} + p \corrinf{\tau_i(1 - \tau_{i-1})}, \qquad i \ge 2.
    \label{eq:ub:J_eqns}
\end{eqnarray}
In the stationary state the time derivatives are zero so \eref{eq:ub:dtJ} defines a conserved current
\begin{equation}
    J_\infty = J_\infty^{(1)} = J_\infty^{(2)} = \ldots
\end{equation}
The bulk current, $J_\infty^{(i)}$, has the usual TASEP hopping term, $p\corrinf{\tau_i(1 - \tau_{i-1})}$.
The additional term $\mu \corrinf{\tau_i}$ arises due to the choice of reference frame.

The two-point correlation $\corrinf{\tau_{i+1}(1 - \tau_i)}$ is computed as
\begin{eqnarray}
    \corrinf{\tau_{i+1}(1-\tau_i)} &= \alpha (1 - \alpha) \sum_{k = 0}^{i-2}\Pjam(k) + \alpha \Pjam(i-1) \nonumber \\
                                &= \alpha(1 - \alpha)\left(1 - \left(\frac{p \alpha}{\mu}\right)^i\right).
    \label{eq:ub:twoPoint}
\end{eqnarray}
The resulting service frame current is
\begin{equation}
    J_\infty = p \alpha(1 - \alpha) + \mu \alpha,
\end{equation}
and is the rate at which particles exit the system.

In terms of the queueing model, the current is the average rate at which high priority customers leave the
queue.  As the total rate at which customers leave the system is $\mu$,\footnote{If the queue was ever empty,
the rate at which customers leave would be less than the service rate, but in the unbounded phase this is not
a concern.} low priority customers leave the queue at rate
\begin{equation}
    \mu - J_\infty = (\mu - p \alpha)(1 - \alpha).
\end{equation}
The constraint $p \alpha < \mu$ (equation \eref{eq:ub:jamConstraint}) ensures that this rate is greater than
zero and low priority customers always receive a share of the service.  For $p \alpha > \mu$, the jam of high
priority customers at the service end becomes unbounded, and low priority customers can no longer reach the
front of the queue to be served.  Thus the low priority current has a second order phase transition at $p
\alpha = \mu$ .

\begin{figure}[ht]
    \centering
        \includegraphics{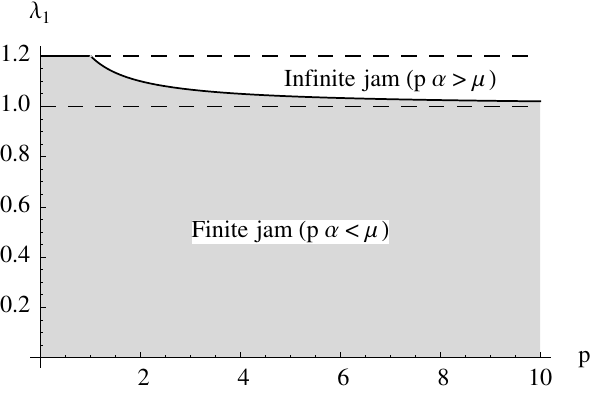}
        \caption{
            Subdivisions of the unbounded phase with $\la = 1.2, \mu = 1$.
        }
        \label{fig:phase_diagram_unbounded}
\end{figure}

The phase transition subdivides the unbounded phase of the PEP.  By fixing values for $\la = \la_1 + \la_2$ and
$\mu$, we can plot illustrative two dimensional phase diagrams with $\la_1$ and $p$ as the axes.  Using
\eref{eq:ub:alphapm} for $\alpha = \alpha_-$, the curve where $p \alpha = \mu$ is given by
\begin{equation}
    \la_1^{(\infty)}(p) =
    \cases{
        \la                                      & $p < \mu$ \\
        \mu\left(1 + \frac{\la - \mu}{p}\right)  & $p \ge \mu$,
    }
\end{equation}
and $p \alpha < \mu$ for $\la_1 < \la_1^{(\infty)}(p)$.  Figure~\ref{fig:phase_diagram_unbounded} shows the
phase diagram for $\la = 1.2, \mu = 1$.
The function $\la_1^{(\infty)}(p)$ is decreasing in $p$ and $\lim_{p \to \infty} \la_1(p) = \mu$.  Therefore, the
transition into the `infinite jam' phase occurs only if $\la_1 > \mu$.  This is marked by the lower dashed
line.  The upper dashed line marks the $\la_1 = \la, \la_2 = 0$ boundary.

\section{The bounded phase}
\label{sec:bounded_phase}

In the queueing theory interpretation, it is the bounded phase of the PEP (with $\la < \mu$) that is of
greatest interest.  In this phase the queue lengths and waiting times remain finite, and we can compare how
the waiting time varies with the overtake rate, $p$, for high and low priority class customers.

The fluctuating lattice length proves a challenge in applying domain wall theory directly to the bounded
phase.  We will take two approaches, each leading to an approximate solution revealing different aspects of
the system.  The first method tells us about the shape and length dependence of the density profiles, while
the second method allows us to calculate customer waiting times.

\subsection{Domain wall ansatz}
\label{sec:boundeddw}

To apply the domain wall ansatz directly in the bounded phase, we consider a general length $n$ configuration
with a length $k$ jam,
\begin{equation}
    \tau_n\ldots\tau_{k+2}01^k = \bm{\tau}01^k.
\end{equation}
The stationary rate equation for $k > 0$, $n > k +1$ is
\begin{eqnarray}
    \fl 0 &= \frac{\dd}{\dd t} P(\bm{\tau}01^k) \nonumber \\
    \fl  &= \la_1 \tau_n P(\tau_{n-1}\ldots\tau_{k+2}01^k) + \la_2 (1 - \tau_n) P(\tau_{n-1}\ldots\tau_{k+2}01^k)
            \nonumber  \\
    \fl  & {} + \sum_{i=k+2}^n p (1 - \tau_i) \tau_{i-1} P(\bm{\tau}01^k|_{(i,i-1)})  + p P(\bm{\tau}101^{k-1})
              + \mu P(\bm{\tau}01^{k+1}) + \mu P(\bm{\tau}01^k0) \nonumber \\
    \fl  & {} - (\la + \mu)P(\bm{\tau}01^k) - \sum_{i=k+2}^n p \tau_i(1 - \tau_{i-1}) P(\bm{\tau}01^k),
    \label{eq:bd:kJamFull}
\end{eqnarray}
and for $k = 0$, $n > 1$
\begin{eqnarray}
    \fl 0 &= \frac{\dd}{\dd t} P(\bm{\tau}0) \nonumber \\
    \fl  &= \la_1 \tau_n P(\tau_{n-1}\ldots\tau_{2}0) + \la_2 (1 - \tau_n) P(\tau_{n-1}\ldots\tau_{2}0)
         \nonumber \\
    \fl  & {} + \sum_{i=2}^n p (1 - \tau_i) \tau_{i-1} P(\bm{\tau}0|_{(i,i-1)})
              + \mu P(\bm{\tau}01) + \mu P(\bm{\tau}00)
         \nonumber \\
    \fl  & {} -(\la + \mu)P(\bm{\tau}0) - \sum_{i=2}^n p \tau_i(1 - \tau_{i-1}) P(\bm{\tau}0).
    \label{eq:bd:0JamFull}
\end{eqnarray}
These are ``bulk" equations, valid when the jam is away from the arrival end of the queue and are of the form
discussed in Section~\ref{sec:DWT}.  To see this, define
\begin{equation}
   P(n, k) = \sum_{\tau_n, \ldots, \tau_{k+2} = 0,1} P(\tau_n\ldots\tau_{k+2}01^k),
\end{equation}
which is the probability of a length $k$ jam in a length $n$ queue.  For $k > 0, n > k + 1$, summing
\eref{eq:bd:kJamFull} and applying the domain wall ansatz \eref{eq:dwprob} gives
\begin{equation}
    \eqalign{
\fl   0 = \frac{\dd}{\dd t} P(n, k) = & \la P(n - 1, k) + \mu P(n + 1, k + 1) + \mu (1 - \alpha) \alpha^k P(n + 1, 0) \\
                                      &  {} + p \alpha P(n, k - 1) - (\la + \mu + p \alpha) P(n, k).
    }
    \label{eq:bd:Pnk} 
\end{equation}
which is exactly \eref{eq:effDW}. And for $k = 0, n > 1$, summing \eref{eq:bd:0JamFull} gives
\begin{equation}
\eqalign{
    \fl 0 &= \frac{\dd}{\dd t} P(n, 0) \\
    \fl   &= \la P(n - 1, 0) + \mu P(n + 1, 1) + \mu (1 - \alpha) P(n+1, 0) - (\la + \mu + p \alpha) P(n, 0).
}
    \label{eq:bd:Pn0}
\end{equation}

The simple domain wall picture breaks down when the jam reaches the arrival end, i.e. for configurations
$01^{n-1}$ or $1^n$.  Our strategy is to find a solution of the bulk equations, without requiring it to
satisfy these boundary equations.  We can hope that this will give an approximation to the true solution.
What we will show is that, within the range of validity, the approximation is very good.

\subsubsection{Length assumption.}
To solve the bulk equations we assume the length dependence factorises as
\begin{equation}
    P(n, k) = P_n \Pjam^*(k),
    \label{eq:bd:lengthAssumption}
\end{equation}
where $P_n$ is the length distribution \eref{eq:lengthDist}.  Then equation \eref{eq:bd:Pnk}, for $k > 0$,
becomes
\begin{equation}
    0 = p \alpha \Pjam^*(k - 1) + \la \Pjam^*(k + 1) + \la (1 - \alpha) \alpha^k \Pjam^*(0)
         - (\la + p \alpha) \Pjam^*(k), \quad
    \label{eq:bd:kjam}
\end{equation}
and equation \eref{eq:bd:Pn0}, for $k = 0$, gives
\begin{equation}
    0 = \la \Pjam^*(1) - (\la \alpha + p \alpha) \Pjam^*(0).
\end{equation}
These have the same form as the unbounded queue domain wall equations, \eref{eq:ub:exactDW},
\eref{eq:ub:zeroDW}, but with $\la$ in place of $\mu$.  Therefore they are solved by
\begin{eqnarray}
\Pjam^*(k) &= \sum_{i=0}^k \left(\frac{p \alpha}{\la}\right)^{k-i} \alpha^i \Pjam^*(0) \nonumber \\ 
&= \frac{p \left(\frac{p \alpha}{\la}\right)^k - \la \alpha^k}{p - \la} \Pjam^*(0).
\label{eq:bd:Pjam}
\end{eqnarray}
The normalisation of $\Pjam^*(k)$ must be independent of $n$.  For the solution to be valid for $n \to \infty$
(as there is no cap on queue length) we must require
\begin{equation}
    \sum_{k = 0}^\infty \Pjam^*(k) = 1,
    \label{eq:bd:jamNorm}
\end{equation}
fixing
\begin{equation}
    \Pjam^*(0) = (1 - \alpha)\left(1 - \frac{p \alpha}{\la}\right),
\end{equation}
subject to the constraint
\begin{equation}
    p \alpha < \la.
    \label{eq:bd:jamConstraint}
\end{equation}
The total probability at each length, $n$, must sum to the length distribution \eref{eq:lengthDist}, that is
\begin{equation}
    \sum_{k = 0}^{n - 1} P(n, k) + P(1^n) = P_n.
\end{equation}
As only $P(1^n)$ is undetermined, we must have that
\begin{eqnarray}
    P(1^n) & = P_n - \sum_{k=0}^{n-1} P(n, k) \nonumber \\
           & = P_n \sum_{k = n}^\infty \Pjam^*(k) \nonumber \\
           & = P_n \frac{p(1 - \alpha)\left(\frac{p \alpha}{\la}\right)^n - \la\left(1 - \frac{p \alpha}{\la}\right) \alpha^n}{p - \la}.
    \label{eq:bd:PAll1s}
\end{eqnarray}
This is analogous to the unbounded queue.  There the probability that the first $n$ sites from the service
end are filled is $\sum_{k=n}^\infty \Pjam(k)$.

To determine $\alpha$ we return to the general $k$-jam equation \eref{eq:bd:kJamFull}\footnote{This is
for  $k>0$, but \eref{eq:bd:0JamFull} for $k=0$ gives the same result.}, and apply the
domain wall ansatz \eref{eq:dwprob} and the length assumption \eref{eq:bd:lengthAssumption}, leaving
\begin{equation}
\eqalign{
    \fl 0 = & \frac{\la_1 \mu}{\la} \tau_n (1 - \alpha) \Pjam^*(k)
                 + \frac{\la_2 \mu}{\la} (1 - \tau_n) \alpha \Pjam^*(k)
                 + p \alpha^2(1 - \alpha) \Pjam^*(k-1) \\
    \fl & {} + \la \alpha(1 - \alpha) \Pjam^*(k+1) + \la (1 - \alpha)^2 \alpha^{k+1} \Pjam^*(0) \\
    \fl & {} - (\la + \mu + p \tau_n)\alpha(1 - \alpha) \Pjam^*(k).
}
    \label{eq:bd:kJamSubbed}
\end{equation}
Multiplying \eref{eq:bd:kjam} by $\alpha(1 - \alpha)$ and subtracting from \eref{eq:bd:kJamSubbed}, we then
consider $\tau_n = 0$ and $\tau_n = 1$ separately.  Both cases reduce to
\begin{equation}
    0 = p \alpha^2 - (p + \mu) \alpha + \frac{\la_1 \mu}{\la},
    \label{eq:bd:alphaQuadratic}
\end{equation}
with solutions
\begin{equation}
    \alpha_\pm = \frac{p + \mu \pm \sqrt{(p+\mu)^2 - 4 p \frac{\la_1 \mu}{\la}}}{2p}.
    \label{eq:bd:alphapm}
\end{equation}
Using the inequalities
\begin{equation}
    \la_1 < \frac{\la_1 \mu}{\la} < \mu,
\end{equation}
(the first inequality holds as the queue is bounded) we see that $0 < \alpha_- < 1$, and $\alpha_+ > 1$
when $p, \la_1, \la_2, \mu > 0$.  Again, we must take $\alpha = \alpha_-$ to have a proper density value.
As for the unbounded queue, in the limit $p \to 0$, $\alpha \to \la_1 / \la$, the expected occupancy of any
site when there is no overtaking.

\subsubsection{Density profile.}
To summarise, we have solved the bulk equations in the domain wall approximation, giving the solution in the form \eref{eq:bd:lengthAssumption}, which with \eref{eq:lengthDist} and \eref{eq:bd:Pjam} results in
\begin{equation}
P(n,k) = \left(1-\frac{\lambda}{\mu}\right) \left(1-\frac{p\alpha}{\lambda}\right)\frac{1-\alpha}{p-\lambda}  \left(\frac{\lambda}{\mu}\right)^n \left(p \left(\frac{p \alpha}{\la}\right)^k - \la \alpha^k\right).
\label{eq:bd:explicitPnk}
\end{equation} 
In general this solution does not satisfy the boundary equations for $k=n, n-1$, i.e. the cases we neglected
were where the jam extends to the length of the queue.  The jam grows with rate $p \alpha$, so the constraint
$p \alpha < \la$ (equation \eref{eq:bd:jamConstraint} arising from the normalisation condition) requires that
the queue length grows faster on average then the jam. Since we have totally neglected the boundary equations,
we expect our approximation to be best when $p\alpha \ll \la$. In fact when $p \to 0$ \eref{eq:bd:explicitPnk}
reduces to
\begin{equation}
    \lim_{p \to 0} P(n, k) = \left(1 - \frac{\la}{\mu}\right)\left(\frac{\la}{\mu}\right)^n
                                \frac{\la_2 \la_1^k}{\la^{k+1}},
\end{equation}
as expected from the exact solution in this case, \eref{eq:pZeroExact}.

\begin{figure}[ht]
    \centering
    \subfigure[
        $\la_1 = 0.7, \la_2 = 0.1, \mu = 1, p = 0.5$, giving $p \alpha < \la$.
    ]
    {
        \includegraphics[width=0.4\textwidth]{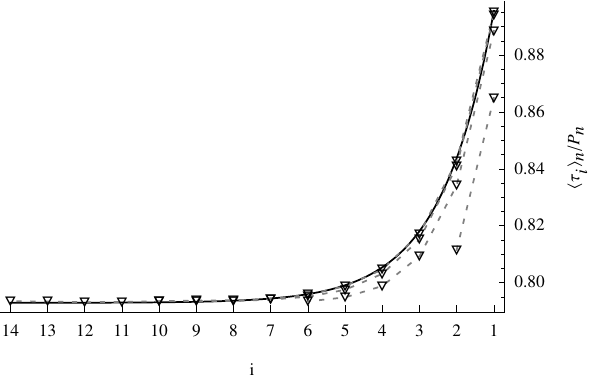}
        \label{fig:lengthdep_dw}
    }
    \qquad
    \subfigure[
        $\la_1 = 0.7, \la_2 = 0.1, \mu = 1, p = 3$, giving $p \alpha > \la$.
    ]
    {
        \includegraphics[width=0.4\textwidth]{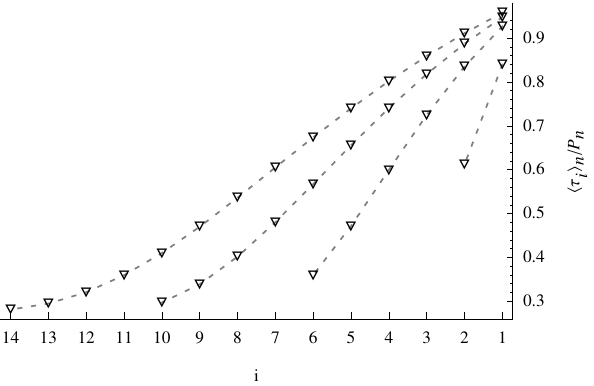}
        \label{fig:lengthdep_nondw}
    }
    \caption{
        Scaled density profiles $\corr{\tau_i}_n / P_n$ for $n = 2, 4, 6, 10, 14$.  Triangle markers
        for simulation results with points for each length $n$ connected by dashed lines.  The analytical
        expression, calculated from \eref{eq:bd:dwdensity}, is plotted as the solid curve in (a).
    }
    \label{fig:lengthdep_density}
\end{figure}

The density at site $i$ in a length $n$ queue, computed from \eref{eq:bd:explicitPnk}, \eref{eq:bd:PAll1s} is
\begin{eqnarray}
    \corr{\tau_i}_n &= \alpha \sum_{k=0}^{i-2}P(n,k) + \sum_{k=i}^{n-1}P(n,k) + P(1^n) \nonumber \\
                    &= \alpha \sum_{k=0}^{i-2}P(n,k) + \sum_{k=i}^{\infty}P(n,k) \nonumber \\
                    &= P_n \left( \alpha + (1 - \alpha) \left( \frac{p \alpha}{\la} \right)^i \right).
    \label{eq:bd:dwdensity}
\end{eqnarray}
Figure~\ref{fig:lengthdep_density} shows length dependent density profiles for $p \alpha < \la$
(Figure~\ref{fig:lengthdep_dw}) and $p \alpha > \la$ (Figure~\ref{fig:lengthdep_nondw}), scaled by dividing
out the length distribution $P_n$.
Triangle markers show simulation results, with points for each length connected by dashed lines.

In Figure~\ref{fig:lengthdep_dw}, the solid curve shows $\corr{\tau_i}_n/P_n$ calculated from
\eref{eq:bd:dwdensity} and \eref{eq:lengthDist}.  We see that as $n$ increases, the simulation results
converge to the calculated profile, and even for $n = 10$ the match is very good.  In
Figure~\ref{fig:lengthdep_nondw}, where the solution \eref{eq:bd:dwdensity} no longer applies, the profiles
become almost linear and are reminiscent of the behaviour of the TASEP on the coexistence line.  In that
phase, the domain wall occurs at all positions with equal probability, resulting, on average, in a linear
profile \cite{SchutzD93}.  In the bounded PEP the evidence suggests that the constraint $p \alpha < \la$ marks
a crossover between the localised jam, for $p \alpha < \la$ and the delocalised jam for $p \alpha > \la$.

\subsection{Aggregate density profile and current}

In this section we take a different approach, working with the rate equations for the one-point functions.  We
start by defining a conserved current for the bounded phase.  To do so, we sum the density at each position
over all lengths, thus aggregating the effect of the length fluctuations.  Define the summed one-point
functions
\begin{equation}
    \corrbar{\tau_i} = \sum_{n=i}^\infty \corr{\tau_i}_n.
    \label{eq:bd:densitySummed}
\end{equation}
Note that as $\corr{\tau_i}_n \le P_n$,
\begin{equation}
    \corrbar{\tau_i} \le \sum_{n=i}^\infty P_n = \left(\frac{\la}{\mu}\right)^i,
    \label{eq:bd:sum_converges}
\end{equation}
so the sum is bounded, and it converges as $\sum_{n=i}^M \corr{\tau_i}_n$ is monotone increasing in $M$.  Thus
$\corrbar{\tau_i}$ is well defined.  Higher order summed correlations are defined similarly, e.g.
\begin{equation}
    \corrbar{\tau_{i+1}(1 - \tau_i)} = \sum_{n=i}^\infty \corr{\tau_{i+1}(1 - \tau_i)}_n.
\end{equation}

Summing the rate equations \eref{eq:rate11} -- \eref{eq:ratein} gives
\begin{eqnarray}
    \fl \frac{\dd}{\dd t} \corrbar{\tau_1} = \la_1 P_0 + \mu \corrbar{\tau_2} + p \corrbar{\tau_2(1 - \tau_1)}
                                            - \mu \corrbar{\tau_1}, \label{eq:bd:summedRate1} \\
    \fl \frac{\dd}{\dd t} \corrbar{\tau_i} = \la_1 P_{i-1} + \mu \corrbar{\tau_{i+1}}
                                + p \corrbar{\tau_{i+1}(1 - \tau_i)}
                                -p \corrbar{\tau_i(1 - \tau_{i-1}} - \mu \corrbar{\tau_i}, \qquad i > 1.
                                \label{eq:bd:summedRatei}
\end{eqnarray}
These can be written as a conservation equation for three currents
\begin{equation}
    \frac{\dd}{\dd t} \corrbar{\tau_i} = J_{\text{ext}}^{(i)} + \overline{J}^{(i+1,i)}
                                                           - \overline{J}^{(i,i-1)}, \qquad i \ge 1,
    \label{eq:bd:3Current}
\end{equation}
where
\begin{eqnarray}
    J_{\text{ext}}^{(i)}   = \la_1 P_{i - 1} \label{eq:bd:Jext}\\
    \overline{J}^{(1,0)}   = \mu \corrbar{\tau_1} \label{eq:bd:Jbar1} \\
    \overline{J}^{(i,i-1)} = \mu \corrbar{\tau_i} + p \corrbar{\tau_i(1-\tau_{i-1})}, \qquad i \ge 2.
    \label{eq:bd:Jbari}
\end{eqnarray}
$\overline{J}^{(i,i-1)}$ is the site-to-site current with a hopping term and frame current term.  But
customers can also step directly into place at the end of the queue, which gives the external current
$J_{\text{ext}}^{(i)}$.

In the stationary distribution the time derivatives are zero, and so \eref{eq:bd:3Current} gives a
recurrence for the site-to-site current
\begin{equation}
    \overline{J}^{(i + 1,i)}  = \overline{J}^{(i,i-1)} - J_{\text{ext}}^{(i)},
\end{equation}
which reduces to
\begin{equation}
    \overline{J}^{(i + 1,i)} = \overline{J}^{(1,0)} - \la_1 \sum_{n=0}^{i-1} P_n.
    \label{eq:bd:JBarRecReduced}
\end{equation}
We have $\lim_{i \to \infty} \overline{J}^{(i+1,i)} = 0$, as both terms on the right hand side of
\eref{eq:bd:Jbari} go to zero.  Therefore \eref{eq:bd:JBarRecReduced} implies that
\begin{equation}
    \overline{J}^{(1,0)} = \la_1 \sum_{n=0}^\infty P_n = \la_1,
\end{equation}
and
\begin{equation}
    \overline{J}^{(i,i-1)} = \la_1 \left(1 - \sum_{n=0}^{i-2} P_n \right)
                           = \la_1 \left(\frac{\la}{\mu}\right)^{i-1}, \qquad i \ge 1.
\end{equation}
Now \eref{eq:bd:Jbar1} tells us that
\begin{equation}
    \corrbar{\tau_1} = \frac{\lambda_1}{\mu}.
    \label{eq:bd:aggcorr1exact}
\end{equation}
This result is exact -- it is the probability that the lattice is at least length one with a particle in site
$1$.

We could have derived this result directly from the notion of the PEP as a queue: high priority customers
leave the system at an average rate $\mu \corrbar{\tau_1}$.  The rate at which high priority customers leave
the system cannot be higher than $\la_1$, the rate they arrive. But as the service capacity exceeds the total
arrival rate ($\la = \la_1 + \la_2$) there is no bottleneck at the server, so high priority
customers\footnote{A corresponding argument applies to low priority customers.} leave at the rate they arrive,
that is $\mu \corrbar{\tau_1} = \la_1$.

The presence of two-point correlations prevent us from calculating exact densities for $i = 2, 3$, etc.  The mean
field method \cite{DerridaDM92,SugdenE07} is the standard way to deal with this, assuming the correlation
between neighbouring sites is small so that the two-point correlations can be approximated as products of the
one-point functions.  But for the PEP, there is a similarity to the unbounded system, which we can exploit to
solve the one-point rate equations.

By the definition \eref{eq:bd:densitySummed}, $\corrbar{\tau_i}$ is the probability
\begin{equation}
    \corrbar{\tau_i} = P(\tau_i = 1, \text{length } n \ge i).
\end{equation}
We can instead work with the conditional probability
\begin{equation}
    \corr{\tau_i | n \ge i} = P(\tau_i = 1 | \text{length } n \ge i);
\end{equation}
the two are related by
\begin{equation}
    \corrbar{\tau_i} = P(\text{length } n \ge i)\corr{\tau_i | n \ge i}
                     = \left( \frac{\la}{\mu} \right)^i \corr{\tau_i | n \ge i}.
    \label{eq:bd:aggregated_conditional}
\end{equation}
Similarly, define $\corr{\tau_i(1 - \tau_{i-1}) | n \ge i}$ through
\begin{equation}
    \corrbar{\tau_i (1 - \tau_{i-1})} = \left( \frac{\la}{\mu} \right)^i \corr{\tau_i (1 - \tau_i)| n \ge i}.
    \label{eq:bd:aggregated_conditional_twopoint}
\end{equation}
Substituting into \eref{eq:bd:Jbar1}, \eref{eq:bd:Jbari} gives
\begin{eqnarray}
    \frac{\la_1 \mu}{\la} & = \mu \corr{\tau_1 | n \ge 1} \nonumber \\
                  & = \mu \corr{\tau_i | n \ge i} + p \corr{\tau_i(1 - \tau_{i-1}) | n \ge i}, \qquad i \ge 2.
    \label{eq:bd:effJ_eqns}
\end{eqnarray}
These have the same form as the current equations for the unbounded queue, \eref{eq:ub:J_eqns}, with an
effective current
\begin{equation}
    \widetilde{J} = \frac{\la_1 \mu}{\la},
\end{equation}
so are solved by the unbounded queue one- and two-point functions \eref{eq:ub:onePoint},
\eref{eq:ub:twoPoint}.  That is
\begin{equation}
    \corr{\tau_i | n \ge i} = \alpha + (1 - \alpha)\left(\frac{p \alpha}{\mu}\right)^i,  \qquad i \ge 1,
    \label{eq:bd:corr_conditional}
\end{equation}
and
\begin{equation}
    \corr{\tau_i(1-\tau_{i-1}) | n \ge i}
        = \alpha(1 - \alpha)\left(1 - \left(\frac{p \alpha}{\mu}\right)^{i-1}\right), \qquad i \ge 2.
    \label{eq:bd:two_corr_conditional}
\end{equation}
To determine $\alpha$, we substitute \eref{eq:bd:corr_conditional}, \eref{eq:bd:two_corr_conditional} into
\eref{eq:bd:effJ_eqns}, and take $i \to \infty$ (assuming $p \alpha < \mu$).  This gives back the quadratic
for $\alpha$ \eref{eq:bd:alphaQuadratic}, so we again must take $\alpha = \alpha_-$, given by
\eref{eq:bd:alphapm}.  Note that $p \alpha < \mu$ if $p, \mu, \la_2 > 0$, so the density profile is always
exponentially decaying.

This solution satisfies the current equation \eref{eq:bd:effJ_eqns}, and therefore the resulting aggregated
one- and two-point functions \eref{eq:bd:aggregated_conditional}, \eref{eq:bd:aggregated_conditional_twopoint}
satisfy the one-point rate equations \eref{eq:bd:summedRate1}, \eref{eq:bd:summedRatei}.  But as it does not
in general satisfy the higher order rate equations it is approximate only.

\begin{figure}[ht]
    \centering
    \subfigure[$\la_1 = 0.1, \la_2 = 0.7, \mu = 1$.]
    {
        \includegraphics[width=0.4\textwidth]{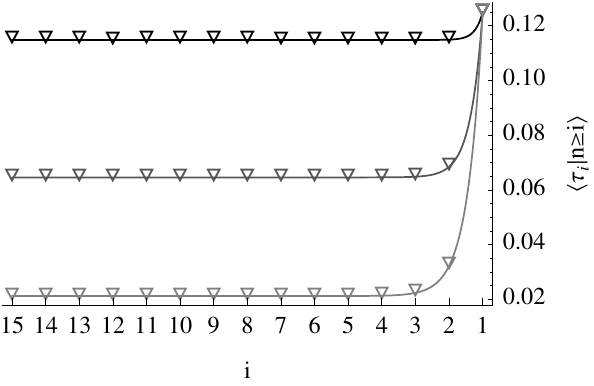}
        \label{fig:bounded_summed_low_density}
    }
    \qquad
    \subfigure[$\la_1 = 0.1, \la_2 = 0.3, \mu = 1$.]
    {
        \includegraphics[width=0.4\textwidth]{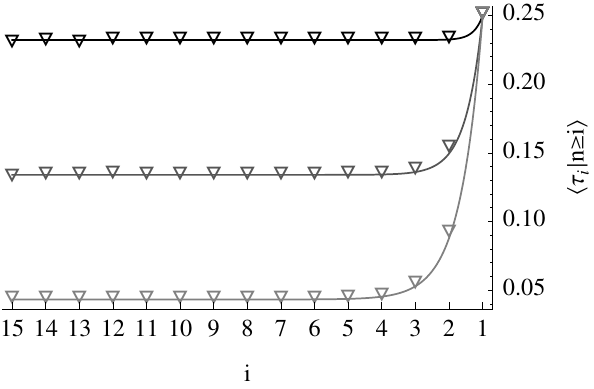}
        \label{fig:bounded_summed_short}
    }

    \subfigure[$\la_1 = 0.7, \la_2 = 0.1, \mu = 1$.]
    {
        \includegraphics{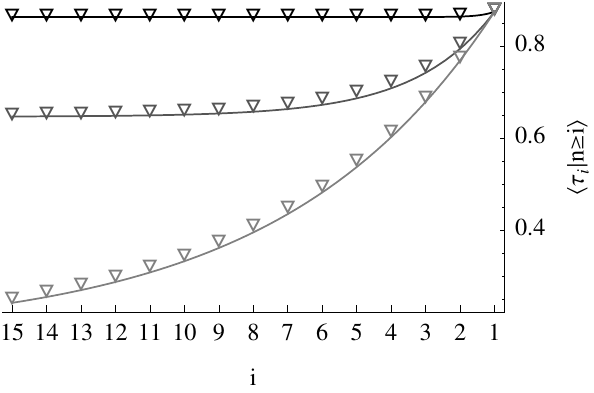}
        \label{fig:bounded_summed_high_density}
    }
    \qquad
    \subfigure[$\la_1 = 0.4, \la_2 = 0.1, \mu = 1$.]
    {
        \includegraphics{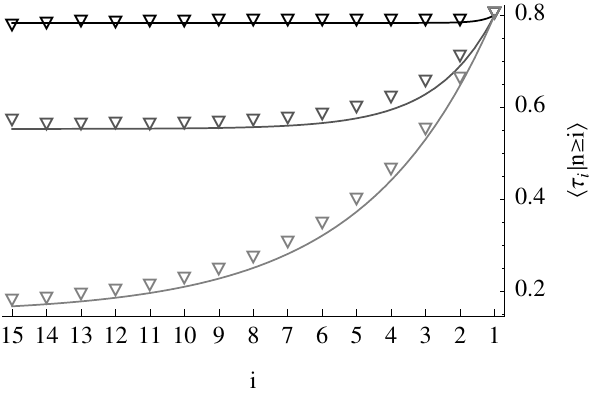}
        \label{fig:bounded_summed_nodw}
    }
    \caption{
        Aggregated density profiles scaled by dividing out $(\la/\mu)^i$.  Triangle markers for simulation results, plotted against
        calculated profile $\corr{\tau_i | n \ge i}$.  For $p = 0.1$ (black), $p = 1$ (mid-gray), $p = 5$
        (light gray).
    }
    \label{fig:density_aggregated}
\end{figure}

Figure~\ref{fig:density_aggregated} compares simulated and calculated  aggregated density profiles,
$\corrbar{\tau_i}$.  We have divided out the length dependence factor $(\la/\mu)^i$ so in fact are plotting
$\corr{\tau_i | n \ge i}$ computed from \eref{eq:bd:corr_conditional}.
We see that at $i = 1$, where the exact value $\corr{\tau_1 | n \ge 1}$ is known, and asymptotically for large
$i$, the simulated and calculated density profiles agree.  At intermediate values of $i$ we see the greatest
discrepancy, indicating that the aggregated domain wall solution is approximate only, although the agreement
is still very good.

Note we could also compare the aggregated profiles with $\corr{\tau_i}_n$ \eref{eq:bd:dwdensity} summed over
$n$.  However, even at position $1$ the summed $\corr{\tau_1}_n$ does not agree with $\corrbar{\tau_1}$
\eref{eq:bd:aggcorr1exact} for which the exact result is known.  The direct application of the domain wall
ansatz in Section~\ref{sec:boundeddw} gave an indication of the length dependence in the system, but the
approach in this section, following from the current conservation equation, is in much better agreement
numerically with simulation results.

\subsection{Waiting times}

We can use the aggregated one-point functions to compute the average number of customers in the queue, and in
turn the average waiting times for both classes of customers.  Switching the order of the sums in
\eref{eq:pepNhigh}, we can write $\overline{N}_1$, the average number of high priority customers, as
\begin{equation}
    \overline{N}_1 = \sum_{i=1}^\infty \sum_{n=i}^\infty \corr{\tau_i}_n
                   = \sum_{i=1}^\infty \corrbar{\tau_i}.
    \label{eq:bd:Nhi}
\end{equation}
Substituting \eref{eq:bd:aggregated_conditional}, \eref{eq:bd:corr_conditional} into \eref{eq:bd:Nhi},
Little's result \eref{eq:littles} gives the average high priority waiting time,
\begin{equation}
    \overline{W}_1 = \frac{1}{\la_1} \overline{N}_1 
                   = \frac{1}{\la_1} \left(
                         \alpha \frac{\la}{\mu - \la} + (1 - \alpha) \frac{p \alpha \la}{\mu^2 - p \alpha \la}
                     \right).
    \label{eq:bd:W1}
\end{equation}
With \eref{eq:pepNlow}, the average low priority waiting time is
\begin{equation}
    \overline{W}_2 = \frac{1}{\la_2} \left(\corr{n} - \overline{N}_1 \right) \nonumber
                   = \frac{1}{\la_2}(1 - \alpha) \left(
                         \frac{\la}{\mu - \la} - \frac{p \alpha \la}{\mu^2 - p \alpha \la}
                     \right).
    \label{eq:bd:W2}
\end{equation}

Though \eref{eq:bd:W1}, \eref{eq:bd:W2} come from an approximate solution, in the $p \to 0$ and
$p \to \infty$ limits they give the correct waiting times.  Taking first the limit $p \to 0$, we find
\begin{equation}
    \lim_{p \to 0} \overline{W}_1 = \lim_{p \to 0} \overline{W}_2 = \frac{1}{\mu - \la}.
    \label{eq:bd:waitPZero}
\end{equation}
With $p = 0$, high and low priority customers are treated identically.  The PEP reduces to a $M/M/1$ queue
with arrival rate $\la$ and service rate $\mu$, for which the average waiting time is as given by
\eref{eq:bd:waitPZero}.

Conversely, if we make the overtake rate infinite, then high priority customers arriving at the
queue will immediately overtake any waiting low priority customers.  In this limit, high priority customers
see an $M/M/1$ queue with arrival rate $\la_1$ and service rate $\mu$, and indeed we find
\begin{equation}
    \lim_{p \to \infty} \overline{W}_1 = \frac{1}{\mu - \la_1}.
\end{equation}
The average waiting time for low priority customers is
\begin{equation}
    \lim_{p \to \infty} \overline{W}_2 = \frac{1}{(1 - \la / \mu)(\mu - \la_1)}.
\end{equation}
This can be found by directly taking the limit, or via the requirement that
$\overline{N}_1 + \overline{N}_2 = \corr{n}$.

\begin{figure}[ht]
    \centering
    \subfigure[$\la_1 = 0.1, \la_2 = 0.7, \mu = 1$]
    {
        \includegraphics{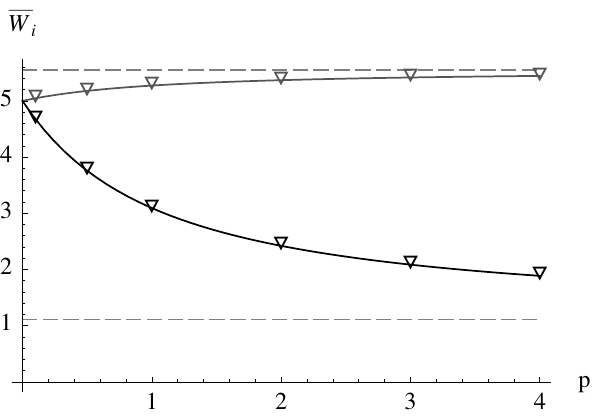}
        \label{fig:waiting_times_low_density}
    }
    \qquad
    \subfigure[$\la_1 = 0.1, \la_2 = 0.3, \mu = 1$]
    {
        \includegraphics{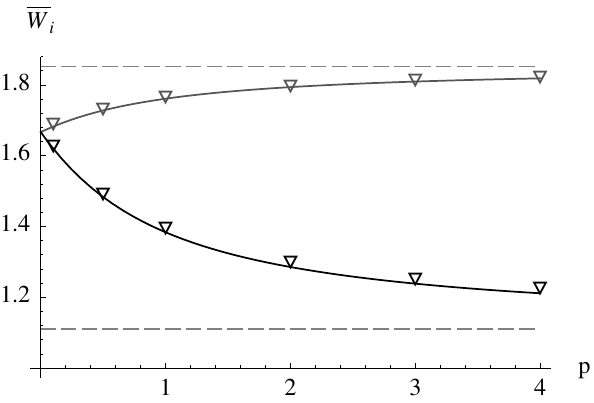}
        \label{fig:waiting_times_short}
    }

    \subfigure[$\la_1 = 0.7, \la_2 = 0.1, \mu = 1$]
    {
        \includegraphics{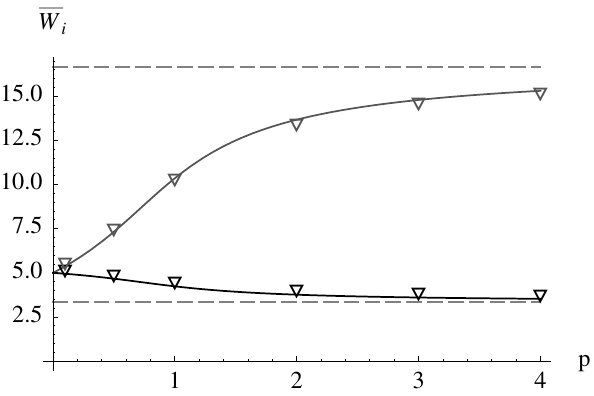}
        \label{fig:waiting_times_high_density}
    }
    \qquad
    \subfigure[$\la_1 = 0.4, \la_2 = 0.1, \mu = 1$]
    {
        \includegraphics{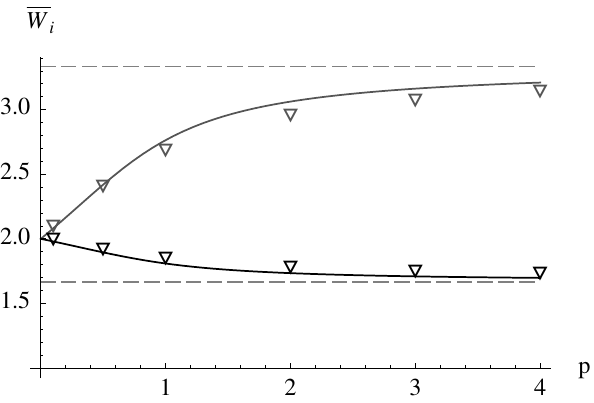}
        \label{fig:waiting_times_nodw}
    }

    \caption{
        Average waiting time for high priority customers (black) and low priority customers (gray)
        plotted against overtake rate $p$.  Dashed lines show the asymptotic values.
    }
    \label{fig:waiting_times}
\end{figure}

Figure~\ref{fig:waiting_times} shows $\overline{W}_1$, $\overline{W}_2$  plotted as a function $p$.  Again we
see good agreement between simulation results and the calculated values.
Increasing $p$ interpolates between a first come first served queue ($p = 0$) and strict prioritisation
according to customer class ($p \to \infty$).  In designing a queueing system, one would choose $p$ to give
the desired high priority waiting time, within the constraints imposed by the asymptotic limits.

The greatest response in $\overline{W}_1$, defined as the maximum value of $|\dd \overline{W}_1 / \dd p|$, occurs at $p = 0$ in Figure~\ref{fig:waiting_times_low_density}, and so there is a strong relative benefit to high priority customers using even small values of $p$. In Figure~\ref{fig:waiting_times_high_density} the value of the parameters give rise to an inflection point at $p > 0$, and hence the largest response in $\overline{W}_1$ occurs at some positive value of $p$. A sufficient
condition for such an inflection point to occur\footnote{This is simpler than trying to solve $\dd \overline{W}_1^2 / \dd p^2 = 0$.}
is $\dd^2 \overline{W}_1 / \dd p^2 |_{p=0} < 0$, which happens if and only if
\begin{equation}
    \frac{\la_2}{\la_1} < \frac{\la_1 + \la_2}{\mu}.
\end{equation}
Then $\dd^2 \overline{W}_1 / \dd p^2$ must change sign as $\lim_{p \to \infty} \dd^2 \overline{W}_1 / \dd p^2
> 0$. For these values of the parameters the benefit to high priority customers of switching on $p$ is relatively small compared to the penalty for low priority customers.

\section{Conclusion}
In this paper we introduce the prioritising exclusion process: a priority queueing model in which high
priority customers are allowed to push ahead in the queue, and thus gain their advantage.  The PEP is
the exclusion process analog of a well studied priority queueing model, the APQ, a connection which is
interesting in itself.  But the PEP also has a natural domain wall structure, which allows domain wall
dynamics to be derived from the microscopic transition rules.  This has recently been achieved for a TASEP
with deterministic bulk motion and stochastic boundary conditions \cite{CividiniHA14}.  In contrast, the PEP
is fully stochastic, but the unique boundary conditions result in the regular behaviour of the jam of high
priority customers, the key descriptor in our domain wall model.

The PEP exhibits a phase transition from a phase with finite expected lattice length, to one with an unbounded
lattice length.  In the unbounded phase, we find the exact solution of the domain wall equations in the
$n \to \infty$ limit, which reveals a further subdivision of this phase into phases with finite or infinite jam length.
We find the condition for an infinite jam, in which case low priority customers will, with probability one,
never get served.  When the jam remains finite we calculate exact stationary density profiles in appropriately
defined local reference frames.

In the bounded phase, domain wall theory does not give exact results but leads to two complementary
approximate solutions. From a direct application of the domain wall ansatz we find that the shape of the
density profile can again be understood in terms of a jam, in this case either localised at the service end or
able to grow and fill the lattice.  In a second approach, a current conservation equation implies that domain
wall theory can be naturally applied to aggregate densities. We thus give very good approximations for these
observables and consequently accurately estimate average customer waiting times. We also give the condition for which
the stochastic overtaking of low priority customers is most effective as a scheduling mechanism.

The behaviour of the jam of high priority customers plays a key role in understanding all phases of the PEP.
There is an interesting analog in the APQ in the notion of \textit{accredited} customers: class $1$
customers with accumulated priority greater than the maximum possible priority of any class $2$ customer
\cite{StanfordTZ14}.  This differs from the definition of a jam, since the last accredited customer may be
followed by an unaccredited class $1$ customer, whereas in the PEP, a jam is always terminated by a  class 2 (low
priority) customer.  Understanding this connection may allow us to compute complete waiting time distributions
for the PEP, as has already been done for the APQ \cite{StanfordTZ14}.

\section*{Acknowledgement}
We thank Peter Taylor for suggesting the PEP to us as well as for discussions, and Alexandre Lazarescu,
Chikashi Arita, Guy Latouce, and Kirone Mallick for discussions and encouragement.  We are grateful to the
Australian Research Council and the ARC Centre of Excellence for Mathematical Frontiers (ACEMS) for financial
support.

\end{document}